\documentstyle[aps]{revtex}
\tighten 
\twocolumn
\draft
\title{Charge injection instability in perfect insulators}

\author{Thomas Christen }

\address{ABB Corporate Research Ltd. \\ CH-5405 Baden-D\"attwil
\\ Switzerland }

\begin{document}

\maketitle
\begin{abstract}
We show that in a macroscopic perfect insulator, charge injection at a
field-enhancing defect is associated with an instability
of the insulating state or with bistability of the insulating and
the charged state.
The effect of a nonlinear carrier mobility is emphasized.
The formation of the charged state is governed by two different processes
with clearly separated time scales. First, due to a fast growth of a
charge-injection mode, a localized charge cloud forms near the injecting defect
(or contact). Charge injection stops when the field enhancement is screened
below criticality. Secondly, the charge slowly redistributes
in the bulk. The linear instability mechanism and the final charged steady state
are discussed for a simple model and for cylindrical and spherical geometries.
The theory explains an experimentally observed increase of the
critical electric field with decreasing size of the injecting contact.
Numerical results are presented for dc and ac biased insulators.
\end{abstract}

\narrowtext

%*********************************************************************
\section{Introduction}
\label{Introduction}
Insulation of dielectrics is limited due to dielectric
breakdown \cite{ODWYER,DISSADO1,ZELLER2}. There exist several different physical mechanisms
which lead to instabilities associated with dielectric breakdown at high electric
fields, e.g., thermal runaway and impact ionization
avalanches. At breakdown, a change from an
insulating to a conducting state occurs, at least in a certain spatial region
and for a certain time. A release of charge carriers is possible from two
different sources. First, carriers can be generated intrinsically 
via a bulk instability, e.g., by ionization of impurities. Secondly,
carriers can enter due to injection at the electrodes.
In this paper, we show that charge injection is associated with an instability, too.
In contrast to bulk instabilities, however, charge injection is
a boundary instability where the unstable mode (charge injection mode)
is localized at the injecting contact.\\
In practice, charge injection in macroscopic insulating bodies occurs at
geometrical defects of the electrodes
where the electric field can be strongly enhanced. Below
we will consider concentric cylindrical and spherical contact
geometries. A small inner electrode has a large electric field and
can serve as a model for a field enhancing defect.
The cylindrical system describes also a
coaxial cable filled with a dielectric medium; a system which is
of obvious technical interest.\\    
Charge injection in dielectrics has been investigated experimentally
for a tip-plate geometry by Hibma and Zeller \cite{HIBMA1}, and has been
modeled by Zeller and Schneider \cite{ZELLER1}
in the limit of an infinitely high mobility edge and by neglecting diffusion.
They describe the mobility edge with a mobility $\mu (E)$ which vanishes for
$E<E_{c}$ but which is very large for $E>E_{c}$, where $E_{c}$ is a critical value of
the electric field ($\approx 10^{7}$V/cm). In this model a space
charge forms near the tip when the local field reaches the mobility edge.
The space charge, in turn, screens the electric field enhancement at the tip
and pins it to the mobility edge (field limiting space charge).
Zeller and collaborators assume a bulk instability at $E_{c}$ associated with
S-shaped negative differential conductivity which forms
the basis of their theory of injection
\cite{ZELLER1,ZELLER3}. Below we show, however, that there is no need for
such an underlying bulk instability. Charge injection turns out to be
an instability by itself.\\
Boggs \cite{BOGGS1,BOGGS2} pointed out the usefulness of the screening
by the injected space charge in ac driven field-grading materials. His model,
however, is based on the concept of conductivity which cannot
lead to a consistent physical description of charge injection.
The theory assumes a conductivity which is only a function of the field and
which does not distinguish between intrinsic and injected charge carriers.
Note that charge injection is a boundary effect,
while conductivity is a bulk quantity associated with intrinsic carriers.
Boggs' approach leads, nonetheless, to qualitatively correct ac results 
in the limit of an infinitely sharp mobility edge and in a certain frequency
regime.\\ 
For the sake of clearness, we will consider a perfect insulator, which is defined
here as a dielectric without intrinsic carriers and
with a constant permittivity $\epsilon $.
Electrons or holes can be present only due to injection at the electrodes.
Without charge injection, a voltage difference at the contacts induces a
charge which is located outside the dielectric in a thin
surface layer (with a thickness of the very short
Debye length) of the metal contacts. We call this locally neutral state of the insulator 
the {\em ideal insulating state}. The electric field in the insulator
is then uniquely determined by the Laplace equation. The electric field at the contacts
is fully determined by the potential differences of the contacts.
Clearly, this situation corresponds to a purely capacitive arrangement of the
contacts in a dielectric medium.\\
A prescription of arbitrary boundary conditions to the electric
field at the contacts (e.g., $E=0$) is more restrictive and
in general implies the presence of a space charge in the dielectric medium. The field is
then determined by the Poisson equation. 
Important work on charge injection \cite{ODWYER,LAMPERT1,LAMPERT2,MOTT1,HARE2} treats
the formation of a space charge in this way as a direct consequence of boundary conditions.
These theories treat the charged state, but they do not consider the stability  of
the locally neutral state.\\
A different approach which is appropriate for metal-semiconductor contacts is to
prescribe a Richardson-Schottky \cite{RICHARDSON1,SCHOTTKY1} or a
Fowler-Nordheim \cite{FOWLER1,MURPHY} current-field characteristic in order to
model thermionic (field)
emission or a tunneling current through the contact barrier, respectively.
In contrast to the well-defined metal-semiconductor micro-contacts manufactured
by a highly developed semiconductor technology, macroscopic metal-insulator contacts 
used in high-voltage devices are not well-defined and can thus not be treated
on a microscopic level.
A description on a hydrodynamic level is then more appropriate.
Therefore, we prescribe phenomenological
boundary conditions to the charge density $\rho $.
Additional boundary conditions to the electric field are
unnecessary. In principle, the parameters occurring in
the boundary conditions should be derived from
microscopic models, provided the physics at the contacts is known.
For homogeneous boundary conditions, it turns out that the locally neutral
state ($\rho \equiv 0$) is always a stationary solution of the problem.
However, we will show that this ideal insulating state can
become unstable against a charge injecting mode or that bistability
of neutral and charged state can occur.\\      
This paper is organized as follows. In the next section, we introduce
a model for the perfect insulator with phenomenological boundary conditions.
In Sect. \ref{Stability}, we investigate the charge injection instability
of the ideal insulating state. The steady state which eventually developes is
discussed in Sect. \ref{charged}. Finally, in Sect \ref{acdriven},
we present numerical results for an ac biased perfect insulator.

\section{The Model}
\label{Model}
We consider a material with electron-hole symmetry
and with an immediate recombination of electrons ($n$) and holes ($p$)
by annihilation.
This means that the mobilities and the diffusion constants
of electrons and holes have equal absolute values, $\mu\equiv -\mu_{n}\equiv \mu_{p}$, and
$D\equiv D_{n}\equiv D_{p}$, respectively. An Einstein relation
is not considered for the present nonequilibrium system, and we assume that
$D$ is a field independent constant. Due to the fast electron-hole recombination, 
the carrier density is equal to the absolute value of the charge density.
The drift current is then simply given by $\mu(E) \vert \rho \vert E$.
Dynamic equations for electrons and holes describing
generation-recombination processes do not appear.
We emphasize that, except for the ac results, all results below
are valid also for unipolar conduction
and are not consequences of the electron-hole symmetry and of the fast recombination.
In particular, the injection instability discussed below occurs for unipolar
conduction.\\
Consider now a perfect insulator in a capacitor of cylindrical
or of spherical symmetry. Metal contacts are attached at the inner and
the outer radius, $r_{1}$ and $r_{2} (\gg r_{1})$, respectively. In the following,
the cylindrical capacitor of length $L_{z}$ and the spherical capacitor
are labeled with $d=1$ and $d=2$, respectively. All quantities depend only
on the radial coordinate, $r$. 
The (radial) current density can be expressed in terms of the charge
density $\rho $ and the (radial) electric field $E$:
\begin{equation}
j = \mu(E) \vert \rho \vert  E - D \partial _{r} \rho \;\;.
\label{eq1}
\end{equation}
We assume a mobility $\mu (E)$ which depends on the field in the form
$\mu(E)E \equiv v \vert E/E_{0} \vert ^{\alpha} {\rm sign}(E)$,
where $\alpha \geq 1$ is a measure of nonlinearity and where $v$ is
a positive velocity. The limit $\alpha \to \infty $ corresponds to the infinitely
sharp mobility edge at $E=E_{0}$ discussed in Ref. \cite{ZELLER1}. Note that already
the case $\alpha =1$ corresponds to a nonlinear current-field
relation since $\rho $ is related to the electric field via
the Poisson equation
\begin{equation}  
\epsilon \nabla \cdot E = \rho \;\; .
\label{eq1a}
\end{equation}
Consequently, a linear dielectric relaxation mode does not exist in the perfect insulator.\\
There are two equivalent
formulations of the dynamics, namely in terms of the Maxwell equation
\begin{equation}  
\epsilon \partial _{t} E = \nabla \times H-j\;\; ,
\label{eq2a}
\end{equation}
which is a dynamic equation for the electric field,
or in terms of the continuity equation,
\begin{equation}  
\partial _{t} \rho = -\nabla \cdot j\;\; ,
\label{eq2b}
\end{equation}
which is a dynamic equation for the charge density. For convenience, we
use below Eq. (\ref{eq2a}) for the numerical simulations and Eq. (\ref{eq2b})
for the analytical discussion.\\
The system is driven electrically via a coupling to an external
electric circuit, which consists here of
a voltage bias $V(t)$ and an ohmic resistor, $R_{ext}$, in series.
The total (radial) current density
$(\nabla \times H)_{r}=J/r^{d}$ is determined by
\begin{equation}  
J = a_{d} \left( V(t)-\int _{r_{1}}^{r_{2}} E\,dr \right)\;\; ,
\label{eq3}
\end{equation}
where $a_{1}=(2\pi L_{z}R_{ext})^{-1}$ and $a_{2}=(4\pi R_{ext})^{-1}$
for the cylindrical and the spherical case, respectively. We mention
that Eq. (\ref{eq3}) gives rise to a strong nonlocality
which can influence qualitatively the
spatio-temporal dynamics of the system \cite{NONLOC}. 
Below, we restrict ourself to the limit case of voltage control, i.e.,
$R_{ext}\to 0$ and to low
frequencies such that inductive effects can be neglected.
An increase of $R_{ext}$ corresponds to forcing
a current which requires the presence of charge and is thus
expected to lower the  stability of the ideal insulating state.
Voltage control can equivalently be expressed
in the form $V(t)=\int _{r_{1}}^{r_{2}}E\,dr$.\\
In order to have a well-defined problem we
specify mixed homogeneous boundary conditions to the charge density
\begin{equation}  
\partial _{r} \rho\vert _{r_{1,2}} \pm \kappa
\rho \vert _{r_{1,2}}=0 \;\; ,
\label{eq2}
\end{equation}
where $\kappa $ is a phenomenological parameter, and where $+$ and $-$ refers
to $r_{1}$ and $r_{2}$, respectively.
Some remarks concerning this boundary condition are in order.
First, a restriction to homogeneous
boundary conditions is not necessary.
An additional inhomogeneity in Eq. (\ref{eq2})
leads to a finite boundary charge.
In this paper, however, we want to show that charge injection occurs
even for homogeneous boundary conditions where a locally neutral
state exists. Secondly, $\kappa $ can depend on the local electric field. Such
nonlinear boundary conditions can lead to instabilities. Below
we show, that even for the linear case an instability occurs,
and we discuss the behavior of the perfect insulator as a function
of $\kappa $. Thirdly, we assume that the charge does
not `wet' the contacts, i.e. $\kappa \leq 0$.
This is reasonable if the microscopic contact potential has the shape of
a barrier. In a purely diffusive system, a `wetting' density
leads to an instability of the uniform state.
For homogeneous Neumann boundary conditions ($\kappa = 0$) which
describe contacts with vanishing diffusion current, the ideal
insulating state in the diffusive regime is marginally stable (gapless
stability spectrum). Indeed, an arbitrary spatially uniform $\rho $
is a solution
in the linear diffusive regime which implies the existence of a
zero mode. For finite negative $\kappa$, the $\rho \equiv 0$ state
is stable in this regime.
In the following section we show that, on the other hand,
an instability of the ideal insulating state
occurs in the drift dominated regime.

\section{Instability of the ideal insulating state}
\label{Stability}
In this and the following section, we consider
a stationary and positive bias voltage applied to the contacts,
$V(t)\equiv V >0$. Obviously, a steady state of the system
is given by $\rho \equiv 0$ and $E=C_{d}/r^{d}$
with $V/C_{1}=\ln (r_{1}/r_{2})$ and $V/C_{2}=r_{1}^{-1}-r_{2}^{-1}$.
This ideal insulating state corresponds to a purely capacitive system.
To test the linear stability of this state, we
seek for the dynamics of a weak perturbation $(\delta E, \delta \rho)\propto \exp(\lambda t)$
which satisfies the boundary conditions (\ref{eq2}). From the
continuity equation (\ref{eq2b}), one finds an eigenvalue equation
for the growth rate $\lambda$ 
\begin{equation}  
\lambda \delta \rho + \frac{1}{r^{d}}\partial _{r}
\left( \frac{(C_{d}/E_{0})^{\alpha}}{r^{d(\alpha -1)}}v\vert
\delta \rho \vert -Dr^{d}\partial _{r} \delta \rho \right) =0\;\; .
\label{eq4}
\end{equation}
An instability of the ideal insulating state occurs if there
exists an eigenvalue $\lambda$ with positive real part, since
the mode $\delta \rho$ associated with such a $\lambda $
grows exponentially in time. A dimensional analysis of Eq.
(\ref{eq4}) leads to a scaling relation for the growth rate,
\begin{equation}  
\lambda = \frac{D}{r_{1}^{2}}f(\Lambda ) 
\label{eq5}
\end{equation}
where $\Lambda =(r_{1}v/D) (E_{1}/E_{0})^{\alpha}$ has the meaning of a
dimensionless control parameter. Here,
$E_{1}=C_{d}/r_{1}^{d}$ is the electric field at the
inner contact. Note that the function $f$ depends still on $d$,
$\alpha$, $\kappa r_{1}$. The dependence on $\kappa r_{2}$ is weak
for $r_{1}\ll r_{2}$ and will be suppressed whenever possible.
The critical field at instability depends
on the various parameters in the form
\begin{equation}
E_{c}=E_{0}\sqrt[\alpha]{\frac{D}{r_{1}v}\Lambda _{c}(\alpha ,d,\kappa r_{1})}
\label{eq6}
\end{equation}
where the function $\Lambda_{c}$ has to be determined from
$f=0$. The eigenvalue $\lambda $ with the largest real part
turns out to be purely real and can be estimated if either the
diffusion current or the drift current dominates. First, if
the drift term can be neglected, Eq. (\ref{eq4}) reduces
to a linear diffusion equation. Consequently, the
eigenfunctions of the stability problem are of diffusion type and are
damped or marginally stable, provided $\kappa \leq 0$.
On the other hand, if the diffusion term can be neglected,
the stability problem reduces to a first order
differential equation. Solving Eq. (\ref{eq4}) for $D=0$ leads to
a positive growth rate
\begin{equation}  
\lambda =
\frac{v}{r_{1}}\left(\frac{E_{1}}{E_{0}} \right)^{\alpha}(d(\alpha -1)
+r_{1} \kappa) \;\; .
\label{eq7}
\end{equation}
associated with a perturbation
\begin{equation}  
\delta \rho (r) \propto
r^{d(\alpha-1)}\exp
\left[ -\left(\frac{r}{r_{1}}\right)^{d\alpha +1}
\frac{d(\alpha -1)+r_{1}\kappa)}{d\alpha +1} \right]\;\; .
\label{eq8}
\end{equation}
Equation (\ref{eq8}) describes the unstable injection mode which is
localized at the inner contact. Obviously, a negative $\kappa$ acts to slow
down the growth of the unstable mode.
In Fig. \ref{fig1} numerical solutions of the
stability problem of the cylindrical case ($d=1$) are shown as a function of $\Lambda $
with $\kappa r_{1} = -0.5 $. For $D/(vr_{1})\to 0$ and $\lambda \geq 0$,
the numerical results are in good accordance
with the approximate analytical results (\ref{eq7}) and (\ref{eq8}).\\
The physical mechanism for the instability can be
understood as a positive feed-back process. Consider a large
electric field at $r=r_{1}$, and assume a negative field
fluctuation, $\delta E <0$ with $\partial _{r}\delta E>-\delta E/r_{1}$, localized
at $r_{1}$.
The Poisson equation implies then a positive charge fluctuation $\delta \rho$
at this contact. Using the linearized Maxwell
equation, $\epsilon \partial _{t}\delta E \approx -\delta j<0$, one
concludes that the negative field fluctuation grows in amplitude, if drift
dominates diffusion. The initial
perturbation is thus amplified which characterizes an instability.
We mention that for $\alpha =1$, the charge
injection mode (\ref{eq8}) has no physical meaning. In this case, the ideal
insulating state is linearly stable, although it is not necessarily
globally stable.\\
For finite $D$, a competition between
the stabilizing diffusion term and the destabilizing drift term leads
to a finite critical value of the control parameter, $\Lambda _{c}$,
or equivalently, to a finite critical field $E_{c}$.
In order to discuss the dependence of $\Lambda _{c}$ in Eq. (\ref{eq6}) on $\kappa r_{1}$,
we solve Eq. (\ref{eq4}) at $\lambda =0$. A solvability condition leads then to
an expression for $\kappa r_{1}$ as a function of
$\Lambda _{c}$ (appendix). The result is plotted in  Fig. \ref{fig2}a for various
values of the nonlinearity parameter
$\alpha $, for $d=1,2$, and for constant radii.
Clearly, $\Lambda _{c}$ vanishes for $\kappa \to 0$.
On the other hand, the locally neutral state becomes more stable as $-\kappa $ increases
due to a decrease of the charge density of a density fluctuation at the injecting contact.\\
In similar way, stability analysis yields the critical field $E_{c}$
as a function of $r_{1}$. We find that the critical field is almost independent
of $r_{1}$ except for small $r_{1}$, where $E_{c}$ becomes large. This behavior
is more pronounced as $\alpha$ increases.
For $\alpha = 3$ and $d=1,2$, the results are shown in Fig. \ref{fig2}b.
For a tip-plate geometry, Hibma and Zeller \cite{HIBMA1} found experimentally 
that the critical field is almost
independent of the tip size in a large range but increases considerably for
very small tip radii. Our theory clearly reproduces this behavior.

\section{The charged steady state }
\label{charged}
The injection of the charge acts to decrease the field enhancement. Consequently,
the growth of the injection mode saturates at a field below
the critical value $E_{c}$. Zeller and Schneider \cite{ZELLER1} observed that in the
infinitely sharp mobility-edge limit, $\alpha \to \infty$, the final state consists of a charged
region with $\rho \propto 1/r$ and $E(r)\approx E_{c}$ for $r_{1}<r<\bar r$, and a
locally neutral region, $\rho \equiv 0$ and $E\propto 1/r^{d}$,
for $\bar r<r<r_{2}$. The outer radius $\bar r$ of the field limiting
space charge is
determined by the continuity of $E(r)$ at $\bar r$ and by the prescribed voltage drop,
$V=\int E\:dr$. One expects for finite $\alpha $ \cite{ZELLER1} and in
the presence of diffusion, that this state decays on a long time scale
and is in fact part of a transient behavior. More concrete, the
charged steady state forms on two clearly separated time scales. On a fast time scale determined
by Eq. (\ref{eq5}), charge is injected such that the electric field drops locally
below the critical field. In a second step, the charge distributes
slowly towards the new stable steady state. The associated time scale is approximately
given by the transit time $\tau_{tr}$ of the domain wall which connects the charged
and the neutral regions. The general discussion of front propagation into
unstable states \cite{SAARLOOS} goes beyond the purpose of this
paper. Here, we give only a rough estimate for a steep charge step
\begin{equation}  
\tau _{tr} \approx \frac{r_{2}}{v}\left(\frac{r_{2}E_{0}}{V}\right)^{\alpha}\;\; .
\label{eq9}
\end{equation}
In particular, we neglected diffusion which acts to slow down the domain wall
velocity and which acts to smear out the domain wall.
Equation (\ref{eq9}) can be obtained by a projection onto the translation
mode of the domain wall and has the simple interpretation that the front 
travels with the drift velocity of the carriers.\\
It should be noted that the slowness of the charge redistribution
indicates a strong dependence on weak perturbations of the homogeneous insulator bulk.
While weak forces are not expected to hinder
the growth of the fast unstable injection mode, the charge redistribution can be
considerably influenced by traps, grain boundaries etc. Therefore an experimental
observation of the slow dynamics and the final state 
discussed below requires a sufficiently clean material. Macroscopic insulating
bodies used in high-voltage devices where the injection instability occurs,
are usually not very clean. Modeling the dynamics of the field limiting space charge
should thus include bulk inhomogeneities.\\
In Fig. \ref{fig3}, we show a numerical simulation of charge injection in the perfect insulator
for $d=1$ and $\alpha =3$. After an increase of the voltage beyond instability threshold,
a charge domain forms. On a long time scale the charge cloud
smears out and relaxes finally towards a $1/r$-like distribution, whereas
the electric field becomes spatially uniform. In order to discuss this final
steady state in the framework of our model,
we first consider the case $D=0$. From $\nabla \cdot j =0$ one finds
a bulk solution
\begin{eqnarray}  
E(r) & = & A_{d}r^{(1-d)/(1+\alpha)} \label{eq10a} \\
\rho (r) & = & \epsilon A_{d}\frac{1+d\alpha}{1+\alpha} r^{-(\alpha+d)/(1+\alpha)}
\label{eq10b}
\end{eqnarray}
with $A_{1}=V/(r_{2}-r_{1})$ and
$A_{2}=V/(r_{2}^{\alpha /(1+\alpha)}-
r_{1}^{\alpha /(1+\alpha)})/(1+\alpha^{-1})$ for the cylindrical
and the spherical case, respectively.
Note the similarity of the electric field distributions for
$d=1$, $\alpha$ arbitrary, and for $\alpha \to \infty$, $d$ arbitrary.
In these cases, $E(r)$ relaxes eventually to a constant value (see also Fig. \ref{fig3}a). 
Clearly, the bulk solution (\ref{eq10b}) does not satisfy the boundary conditions. 
For a small diffusion constant, the solution is expected to be
changed considerably only in a boundary layer near the contacts. We find that
the solution deep in the bulk far away from the contacts is only weakly disturbed by
a small diffusion constant. For the cylindrical geometry ($d=1$) and for
$V<r_{2}E_{0}$, the bulk solution reads in leading order of $D$ 
\begin{equation}
E(r)=\frac{V}{r_{2}}(1-\frac{D}{\alpha v r}(\frac{r_{2}E_{0}}{V})^{\alpha})\;\;.
\label{eq11}
\end{equation}
For $d=\alpha = 1$, Eq. (\ref{eq11}) is the exact bulk solution.\\
For $\alpha = 1$ we do not find a linear instability
of the ideal insulating state neither numerically nor with the
analytical approximation (\ref{eq7}). However, we
conjecture bistability of neutral
and charged state for $V>(D/\mu)\ln (r_{2}/r_{1})$ and a loss
of global stability at a certain field. Below, this
conjecture will be confirmed with the help of a simulation.
A detailed investigation of this case, however, will be published
elsewhere.\\
Assuming $r_{2}\gg r_{1}$, Eqs. (\ref{eq10a}) and (\ref{eq10b})
yield the current 
\begin{equation}
I_{d}=b_{d}\epsilon v E_{0}\left( \frac{V}{r_{2}E_{0}}\right)^{\alpha +1}\;\;,
\label{eq12}
\end{equation}
where $b_{1}=2\pi L_{z}$ and
$b_{2}=4\pi r_{2}\alpha ^{1+\alpha}(1+2\alpha)/(1+\alpha)^{2+\alpha}$
for the cylindrical and the spherical case, respectively.
Just above instability, the current is finite, though it is small
(of ${\cal O}((r_{1}/r_{2})^{\alpha+1})$).
We mention that for $d=2$, $\alpha =1$ and without diffusion,
the $1/\sqrt{r}$ behavior of the electric field (\ref{eq10a}) is discussed by
Lampert and collaborators \cite{LAMPERT1,LAMPERT2}. Neglecting diffusion, they obtain a
boundary layer due to an $E=0$ boundary condition at the contact. 
One recovers from Eq. (\ref{eq12}) in this case the current-voltage characteristic
of the perfect insulator in a spherical conductor,
$I=(3/2)\pi \epsilon \mu V^{2}/r_{2}$. A detailed discussion of the steady state,
e.g. in the presence of intrinsic carriers, can be found also in Refs.
\cite{LAMPERT1,LAMPERT2}. 

\section{The ac driven insulator}
\label{acdriven}
The localized field limiting space charge which forms
at instability
can be observed in ac experiments \cite{ZELLER2,HIBMA1}.
Since $\lambda$ depends exponentially on $\alpha $, the injection mode grows
infinitely fast at $E_{c}=E_{0}$ in the limit of infinitely
sharp mobility edge ($\alpha \to \infty$).
Consequently, the electric field saturates immediately at $E\equiv E_{c}$ due
to sceening. On the other hand, the characteristic
time of the charge redistribution, Eq. (\ref{eq9}), diverges, provided
$E_{c}r_{2}>V$. For $E_{c}r_{2}<V$ the whole bulk is charged up.\\
In the following, we consider a periodic voltage which vanishes for
$t<0$ and which, for positive $t$, is given by
$V(t)=\hat V\sin (\omega t)$. The frequency $\omega=2\pi /T$ obeys
$\omega \tau _{tr}\gg 1 \gg \omega /\lambda $,
where $\lambda $ is the steady-state stability eigenvalue
for an electric field equal to the amplitude of the
electric field oscillation. Since $\lambda ^{-1}\ll 1$ms and the transit
time $\tau _{tr}$ is of the order
of hours \cite{HIBMA1}, reasonable frequencies are in the range of
$10^{-1}-10^{3}$Hz.\\   
Typical solutions are shown in Fig. \ref{fig4} for
the cylindrical geometry and for various values of $\alpha$. The thin solid line
represents the reference electric field $E_{1}/E_{0}$ at $r_{1}$ of the ideal insulating
state with a purely capacitive response. The other curves represent numerical
simulations of the ac response in the presence of injection. For fields
below instability threshold, $E_{1}<E_{c}$, the sample remains locally
neutral. An increase of the field
beyond threshold leads to the injection of charge in such a way that
the local electric field at the inner contact is saturated slightly below
the critical field. Negative and positive charge is periodically injected
for $\alpha = 3$ (dotted curve). Clearly, due to the electron-hole symmetry the
solutions are symmetric with respect to inversion of the sign of the amplitude.
The critical field is about $2E_{0}$, where the field drops fast and saturates below
$E_{0}$. This discontinuous transition from the neutral state
to the charged state indicates also bistability.
In the limit of a sharp mobility edge (solid curve; $\alpha=51$),
the neutral state decays immediately at $E_{1}=E_{0}$ and the electric field
oscillates between $\pm E_{0}$, as expected.\\
On the other hand, for
$\alpha = 1$ the insulating state $\rho \equiv 0$ is linearly stable even
for large amplitudes $\hat V$.
For certain initial conditions or in the presence of additional current noise, 
however, we find also periodic solutions where only
positive or only negative charge is injected.
An example for positive charge injection is given by the dashed curve.
Charge injection occurs
at a large field amplitude $E_{1}\approx 9 E_{0}$. Once charge has been injected,
a part of it remains in the sample and decreases the value of the electric
field (dashed curve) compared to the chargeless case (thin solid curve).
During the negative half-cycle, the electric stress is thus enhanced.
Due to the presence of this charge, a further injection occurs at a much lower field in
the second cycle of the oscillation.\\
For initial conditions with reversed sign and
$\hat V\to -\hat V$ we find injection of charge with a different sign.
We conclude that there are (at least) three different attractors indicating
bistability in the stationary case. Consequently, even in a system with electron-hole symmetry,
rectification is possible due to dynamical symmetry breaking of charge injection.

\section{Conclusion}
\label{Conclusion}
We have investigated charge injection in a macroscopic and perfect
insulator and for cylindrical
and spherical geometries of the electrodes. The injecting
metal-insulator contacts are modeled on a hydrodynamic
level with boundary conditions for the charge density.
We showed that, depending on the nonlinearity of the mobility,
the ideal insulating state $\rho \equiv 0$ is unstable against a
charge injection mode at a critical field $E_{c}$. Former theories
on charge injection assume either an intrinsic instability \cite{ZELLER3}
or force injection directly by boundary conditions incompatible with
a charge neutral state \cite{LAMPERT1,LAMPERT2}. These theories
cannot predict a finite critical field for the charge injection instability.
For a macroscopic metal-insulator contact which
is usually not well-defined, phenomenological boundary conditions to the
charge density (which serves as an order-parameter field) is appropriate.   
Our theory predicts not only a critical
injection field, $E_{c}$, but reproduces also the experimentally
observed increase of $E_{c}$ with decreasing radius of the injecting contact
\cite{HIBMA1}.\\
For a constant mobility ($\alpha =1$), the ideal insulating state is
linearly stable. But numerical ac simulations which are adiabatic
on the fast time scale show a decay of the neutral state
to a charged state.
From this we concluded bistability and a loss of global
stability of the ideal insulating state.\\
For $\alpha >1$, the time evolution from the insulating to the charged state occurs
on two clearly separated time scales. On a short time scale, the injection mode
grows at the instability and screens the electric field enhancement.
A localized charge cloud forms
near the contact. On a long time scale this charge redistributes over
the whole sample. The localized field limiting space charge can be investigated
with the help of an ac-bias which oscillates with a characteristic time lying between
the just mentioned time scales.\\
Furthermore, we discussed the charged steady
state for small diffusion constants. The current-voltage characteristics of these solutions
are determined by the bulk properties and are thus equivalent to earlier results
by Lampert and coworkers \cite{LAMPERT1,LAMPERT2} in the perfect insulator limit.\\
Future work should address the following problems.
First, the transport model must be refined to include additional
physical effects such as the influence of intrinsic carriers, boundary states,
traps and impurity-band conduction, surface potential decay, and (bi-)polarons.
The effect of traps
enters already by part via the diffusion constant. A small intrinsic
carrier density is expected to increase the stability of the neutral state
due to a finite dielectric relaxation mode. Furthermore,
electron-hole symmetry is not very realistic and one expects rectification
in ac experiments \cite{ARALDIT}.\\
Of interest is also the inclusion of heat transport
and the influence of the temperature, and of mechanical stresses.
Another important task is the determination of the parameters appearing in
the boundary conditions for the charge density from microscopic models.
This is reasonable, however, only for physically well-defined contacts which is
usually not the case for typical macroscopic metal-insulator contacts.\\     
The following interesting problem concerns the case of a constant mobility without
linear instability. Injection should then be associated with  
a nucleation of the charged phase at the contact \cite{CHRISTEN1}, probably via a
critical droplet with a the shape similar to the injection mode (\ref{eq8}). Such an
injection mechanism
is also possible for bistability at $\alpha >1$ in the region where
the ideal insulating state is `supersaturated'.\\
Finally, in order to quantitatively compare theoretical with experimental
results, one should investigate geometries different
from cylindrical and spherical symmetry
as, e.g., a tip-plate arrangement \cite{ZELLER1,SCHNEIDER1,HARE1}.
In contrast to the simple finite difference methods used in the simulations
of this work, finite element methods are more appropriate
to simulate charge injection in such real two- or three-dimensional geometries.\\

{\em Acknowledgments}
I am grateful to J. Rhyner for many valuable discussions and for a careful
reading of the manuscript. I thank F. Stucki for drawing my attention to
the relevant literature.

\section*{appendix}
We calculate the relation between the critical value $\Lambda _{c}$ and
the boundary-condition parameter $\kappa$ by solving Eq. (\ref{eq4})  
with the boundary condition (\ref{eq2}) for $\delta \rho$
at $\lambda =0$. We can assume $\delta \rho >0$. Integration of
Eq. (\ref{eq4}) leads to
\begin{displaymath}
\frac{\Lambda _{c}}{r^{d\alpha}}
\delta \rho -\partial _{r}\delta \rho =\frac{A_{2}}{r^{d}}\;\;.
\end{displaymath}
with a constant $A_{2}$. The general solution of this equation is
\begin{displaymath}
\delta \rho = \left( A_{1} -A_{2}\int^{r}\frac{d\tilde r}{{\tilde r}^{d}}
\exp (\frac{\Lambda _{c}\tilde r ^{1-\alpha d}}{\alpha d -1} ) \right)
\exp (\frac{\Lambda _{c}\tilde r ^{1-\alpha d}}{1-\alpha d } ) \;\;.
\end{displaymath}
Applying the boundary conditions (\ref{eq2}) to this function
yields two linear equations for the constants $A_{1}$ and $A_{2}$.
The existence of a non-trivial solution requires the vanishing of
the determinant associated with these equations. This condition
can be written in the form $a\kappa ^{2} +b\kappa +c=0$ with constants
$a,b,c$ depending on $\Lambda$. The solution for negative $\kappa$
defines the stability boundary plotted
in Fig. \ref{fig1}a. In a similar way one calculates the dependence
of $E_{c}$ on $r_{1}$ which is
shown in Fig. \ref{fig1}b.
  
%%%%%%%%%%%%%%%%%%%%%%%%%%%%%%%%%%%%%%%%%%%%%%%%%%%%%%%%%%%%%%%%%%%%%%%%

%%%%%%%%%%%%%%%%%%%%%%%%%%%%%%%%%%%%%%%%%%%%%%%%%%%%%%%%%%%%%%%%%%%%%
\newpage

\begin{figure}[htb]
\caption{a) Largest eigenvalue $\lambda$
of the stability problem as a function
of the control parameter $\Lambda$ ($\kappa r_{1}=-0.5$ $d=1$, $\alpha =3$).
b) Eigenfunctions of the stability problem. The solid curve represents the marginal
charge-injection mode ($\lambda =0$) at $\Lambda =\Lambda _{c}$.
Modes are more localized at the inner contact as
$\Lambda$ increases.}
\label{fig1}
\end{figure}

\begin{figure}[htb]
\caption{a): Critical values $\Lambda _{c}$ as a function of
$ -\kappa r_{1}$, for
cylindrical (solid) and spherical (dashed) geometries  ($r_{1}/r_{2}=0.01$).
Different curves with decreasing stability threshold 
belong to $\alpha =1, 3$, and $15$. b)
Critical value of the electric field at the inner electrode
as a function of the size of the injecting electrode
($\alpha = 3$; solid: $d=1$, dashed: $d=2$).}
\label{fig2}
\end{figure}

\begin{figure}[htb]
\caption{Evolution of a) the electric field distribution and b) the charge density
distribution beyond stability threshold. The localized injection mode
(dotted curve) grows up to a certain amplitude. The domain wall
moves into the bulk (dashed-dotted curves) until the final steady state
with a uniform field (solid curve) is reached ($d=1$, $\alpha =3$).}
\label{fig3}
\end{figure}
\begin{figure}[htb]
\caption{Time dependence of the ac field at the inner electrode
for $\alpha=1$ (dashed), $\alpha = 3$ (dotted)
and $\alpha = 51$ (solid); $vr_{1}/D=0.1$. Solid thin curve: field without injection.}
\label{fig4}
\end{figure}


\begin{thebibliography}{9}
\bibitem{ODWYER} J. J. O'Dwyer, `The theory of electrical conduction and
breakdown in solid dielectrics' (Clarendon Press, 1973).
\bibitem{DISSADO1} L. A. Dissado and J. C. Fothergill, `Electrical
degradation and breakdown in polymers' (Peter Peregrinus Ltd., London, 1992).
\bibitem{ZELLER2} H. R. Zeller et al., `The physics of electrical breakdown
and pre-breakdown in solid dielectrics' in {\em Festk\"orperprobleme 27}
              (Vieweg 1987) p. 223.
\bibitem{HIBMA1} T. Hibma and H. R. Zeller,
           J. Appl. Phys. {\bf 59}, 1614 (1986).
\bibitem{ZELLER1} H. R. Zeller and W. R. Schneider,
           J. Appl. Phys. {\bf 56}, 455 (1984).
\bibitem{ZELLER3} H. R. Zeller,
           IEEE Trans. EI {\bf 22}, 115 (1987).
\bibitem{BOGGS1} S. A. Boggs, IEEE Trans. EI {\bf 28}, 365 (1993).
\bibitem{BOGGS2} S. A. Boggs, IEEE Trans. DEI {\bf 2}, 97 (1995).
\bibitem{LAMPERT1} M. A. Lampert and P. Mark, `Current injection in solids',
(Academic Press, New York, London 1970).
\bibitem{LAMPERT2} M. A. Lampert and R. B. Schilling, in `Semiconductors
and Semimetals', edited by R. K. Willardson and A. C. Beer, Vol 6 
(Academic Press, New York, London, 1970) p. 1.
\bibitem{MOTT1} N. F. Mott and R. W. Gurney,
              `Electronic Processes in Ionic Crystals' (Clarendon Press, Oxford 1940).
\bibitem{HARE2} R. W. Hare, R. M. Hill, and C. J. Budd,
           J. Phys. D: Appl. Phys. {\bf 26}, 1084 (1993), and refs. cited therein.
\bibitem{RICHARDSON1} O. W. Richardson, 'The Emission of Electricity
          from Hot Bodies', (Longmans Geen and Co., London, 1921).
\bibitem{SCHOTTKY1} W. Schottky, Phys. Z. {\bf 15}, 872 (1914).
\bibitem{FOWLER1} R. H. Fowler and L. Nordheim,
           Proc. R. Soc. A {\bf 119}, 173 (1928).
\bibitem{MURPHY} E. L. Murphy and R. H. Good, Phys. Rev. {\bf 102}, 1464 (1956).
\bibitem{NONLOC} F. J. Elmer, Phys. Rev. A {\bf 41}, 4174 (1990).
\bibitem{SAARLOOS} W. van Saarloos, Phys. Rev. A {\bf 37}, 211 (1988); Phys. Rev. A
                {\bf 39}, 6367 (1989).
\bibitem{ARALDIT} While epoxy resin shows a symmetric response,
                 polyethylene is rectifying; T. Baumann et al,
                 internal report, BBC Research Center, Baden, Switzerland,
                 (1987). 
\bibitem{CHRISTEN1} T. Christen, Europhys. Lett. 31, 181 (1995). 
\bibitem{SCHNEIDER1} W. R. Schneider, unpublished.
\bibitem{HARE1} R. W. Hare and R. M. Hill,
           J. Phys. D: Appl. Phys. {\bf 24}, 398 (1991).

 

\end{thebibliography}
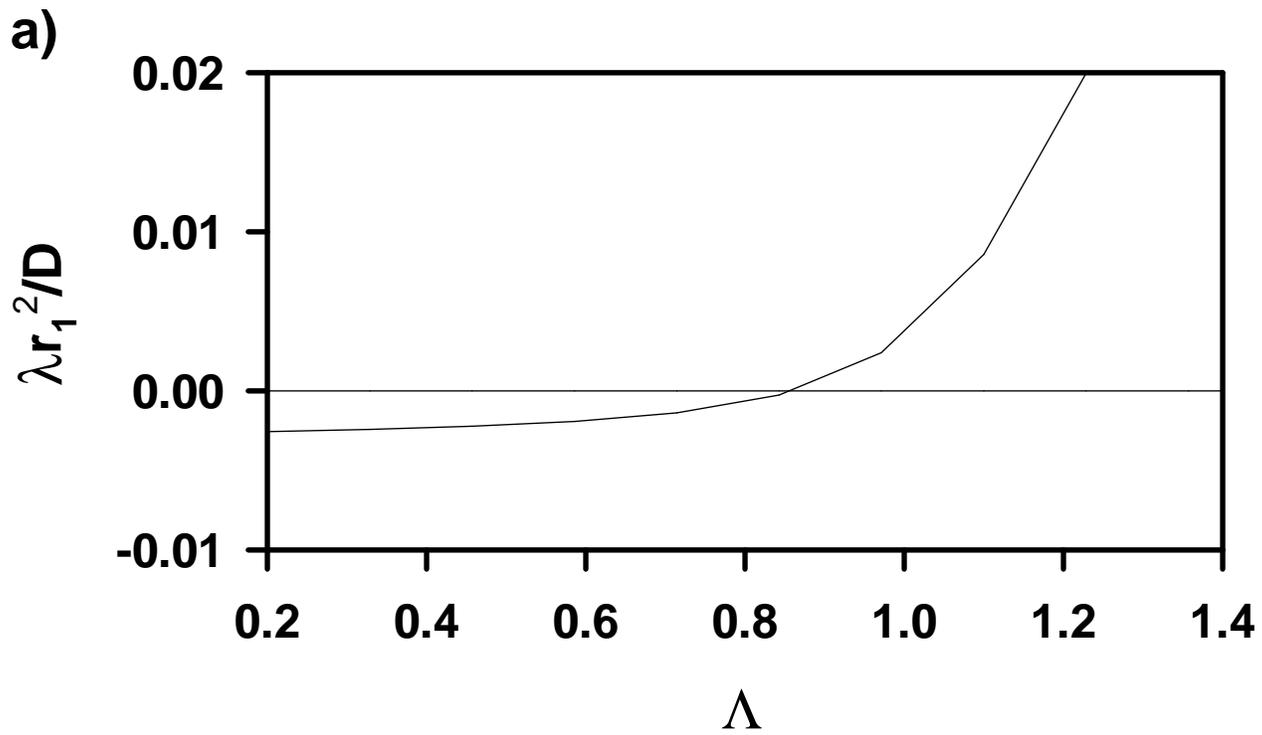
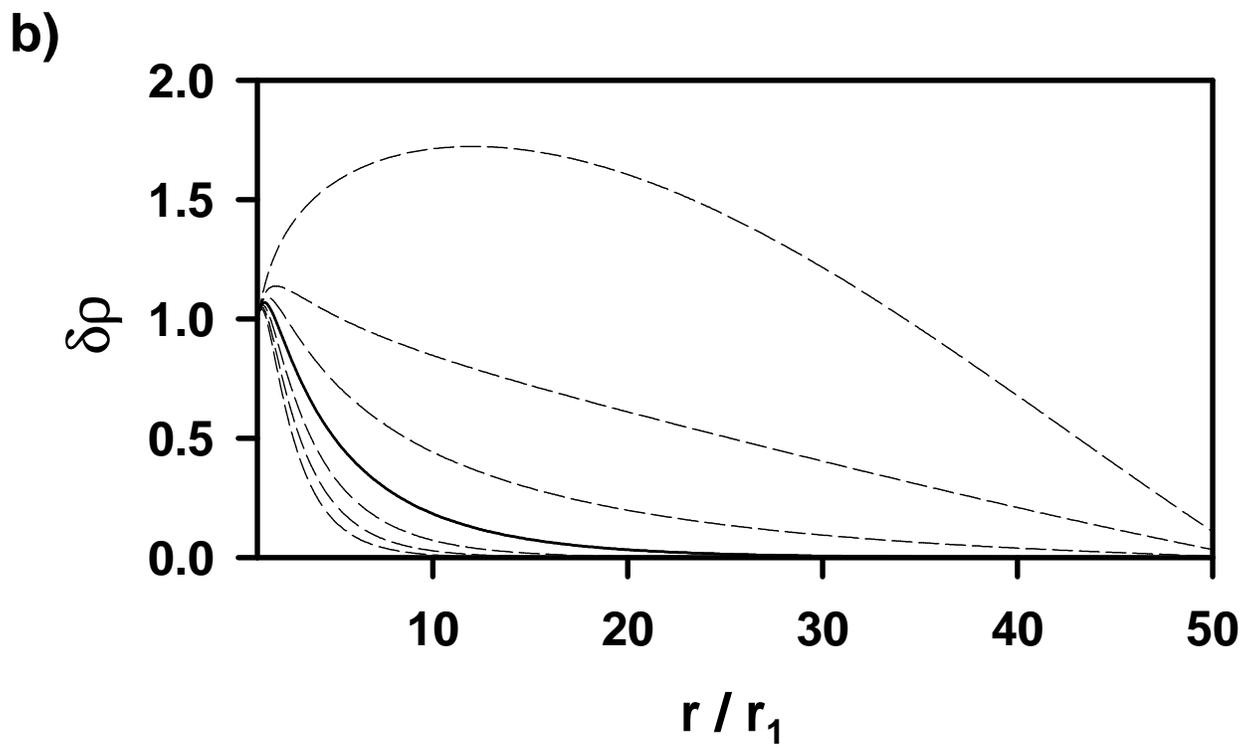
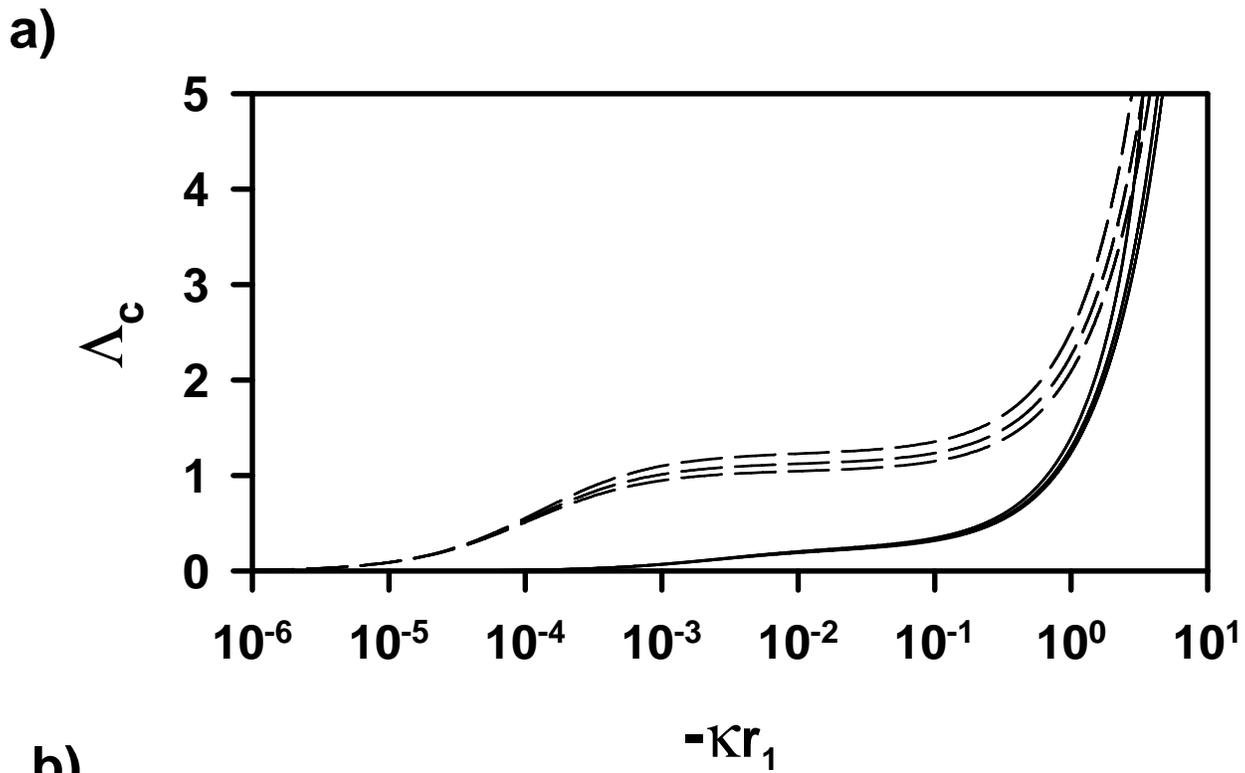
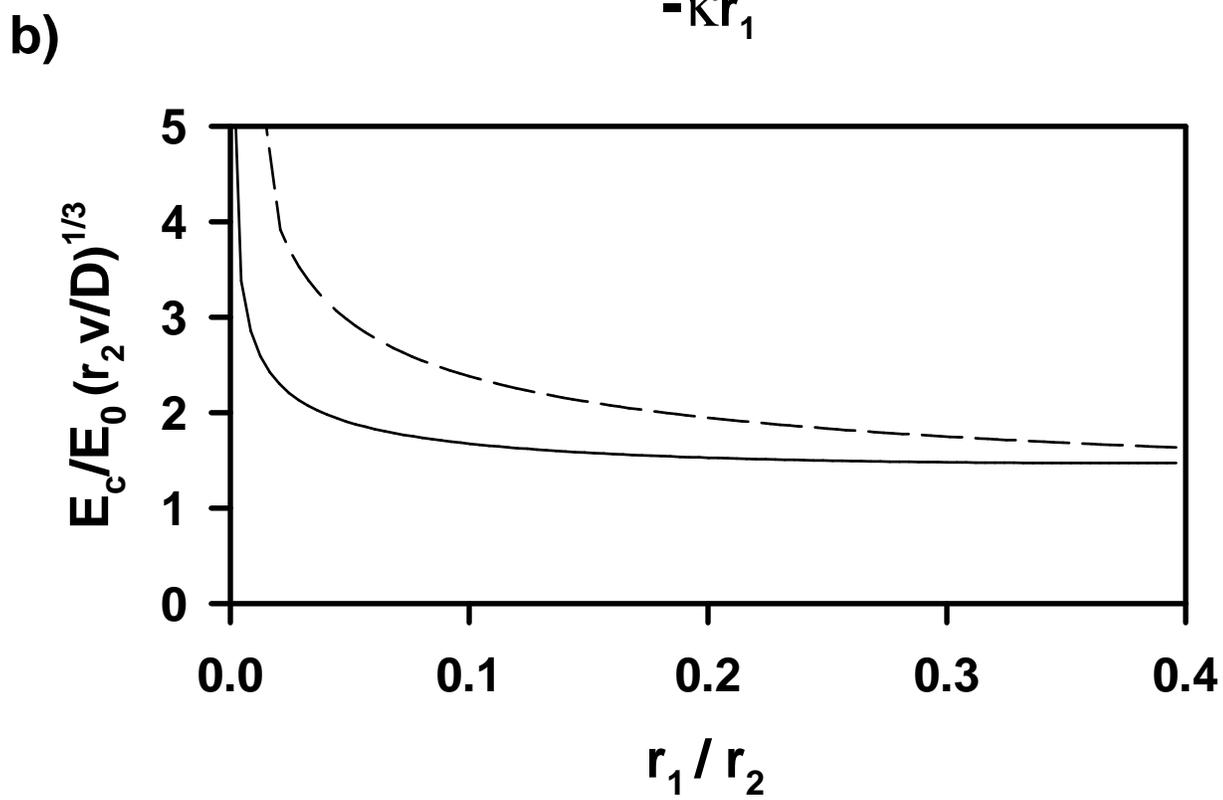
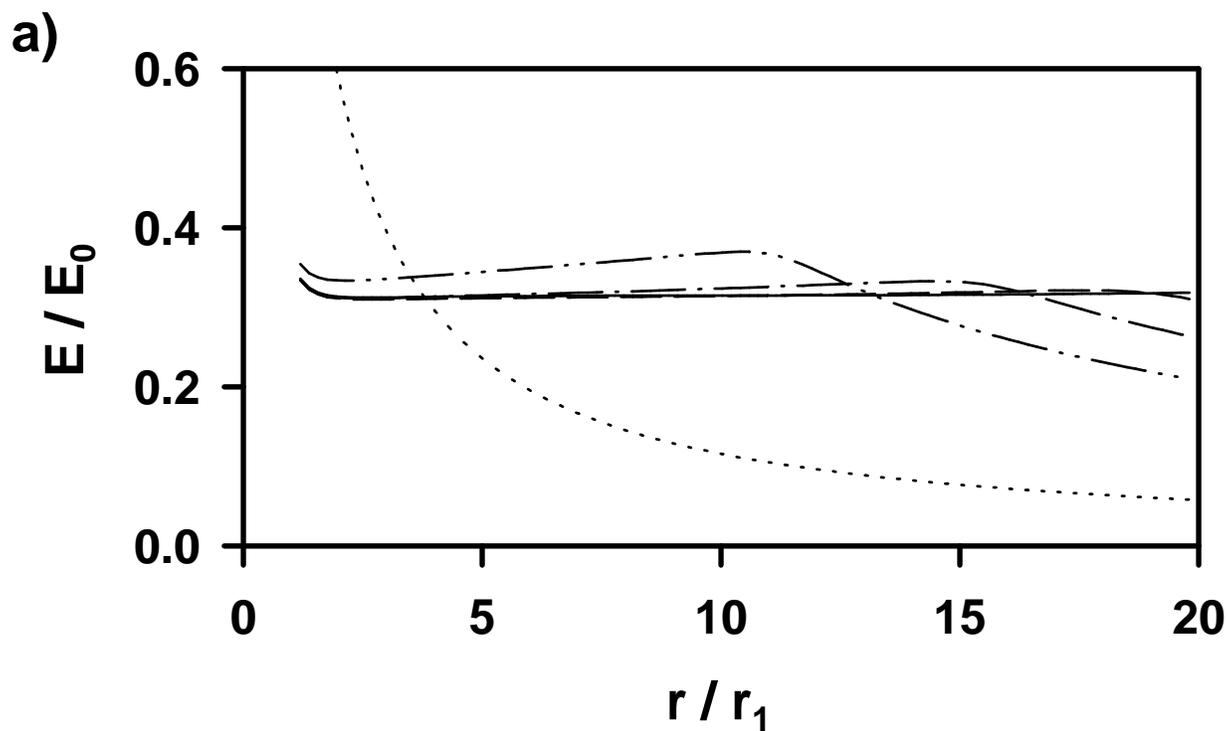
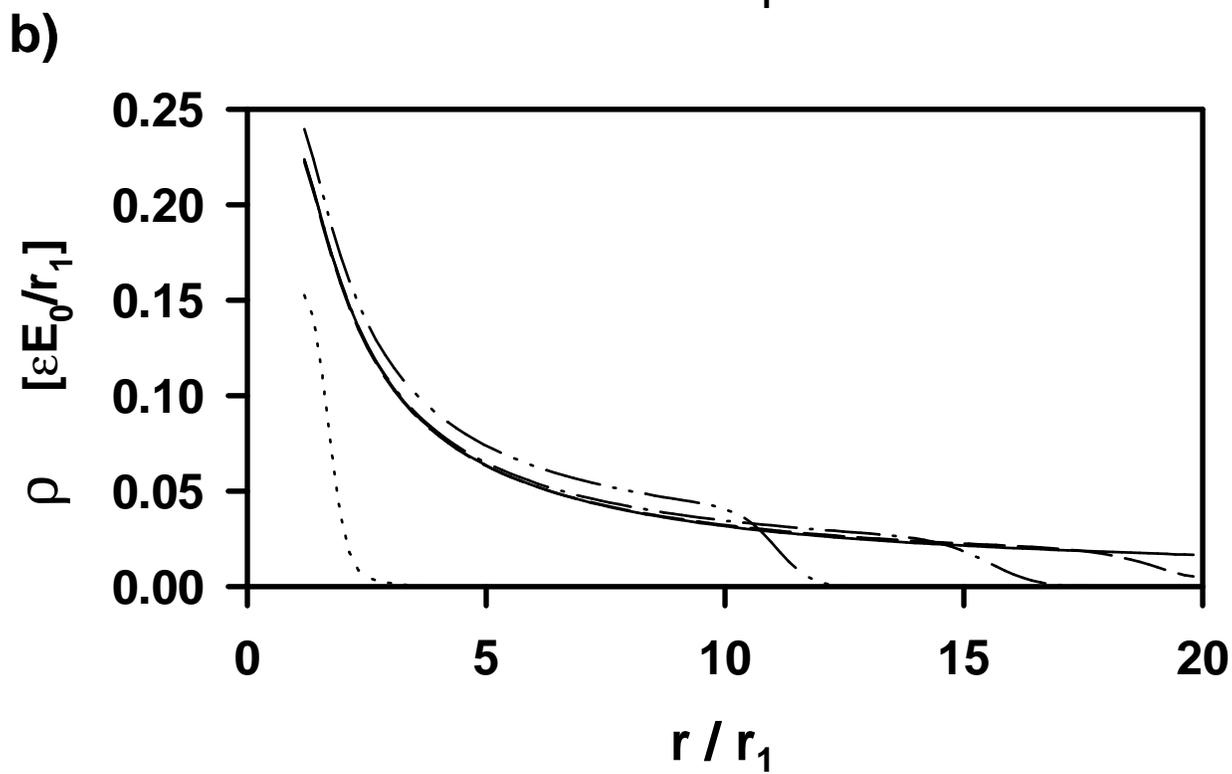
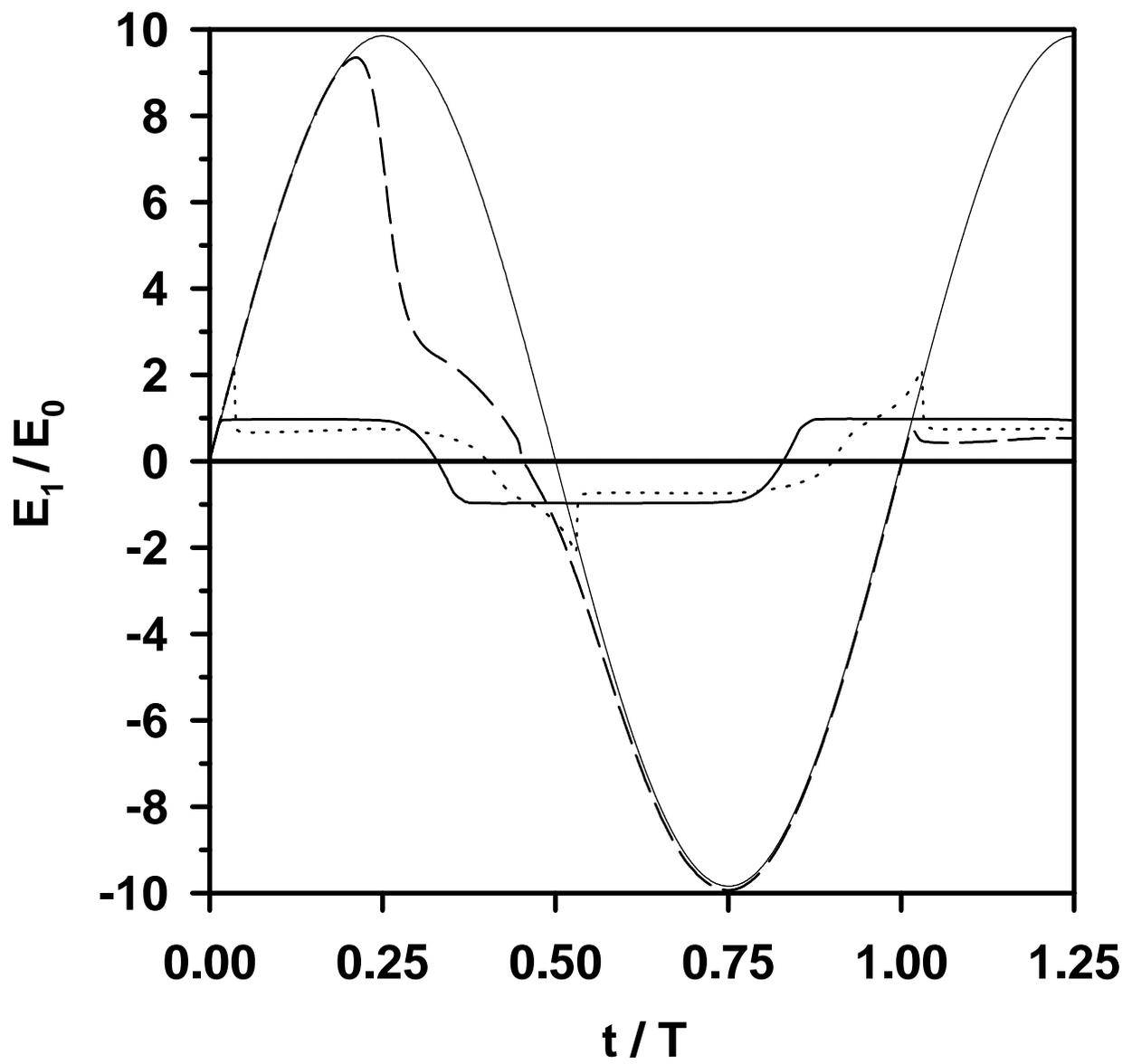
\end{document}